\documentclass[12pt,preprint]{aastex}

\shorttitle{Dust or C-stars?} \shortauthors{Dottori et al.}

\begin{document}

\title{Separating C-stars from Dust in the Central Region of the Seyfert\,2 Galaxy NGC\,1241}

\author{Horacio Dottori\altaffilmark{1}, Rub\'en J.
D\'{\i}az\altaffilmark{2}, Gustavo Carranza\altaffilmark{2},
Sebasti\'an L\'{\i}pari\altaffilmark{2} \& Jo\~ao Santos
Jr.\altaffilmark{3}}

\altaffiltext{1}{Instituto de F\'{\i}sica -- Universidade Federal
do Rio Grande do Sul, Porto Alegre, RS, Brazil. \\e-mail:
dottori@if.ufrgs.br} \altaffiltext{2}{Observatorio Astron\'omico
de C\'ordoba, UNC, Laprida 854, 5000 C\'ordoba, and CONICET,
 Argentina}
\altaffiltext{3}{Departamento de Fisica, ICEx, UFMG, CP 702,
30123-970, Belo Horizonte, MG, Brazil}
\begin{abstract}

The Sy\,2 galaxy NGC\,1241 presents a 1.5\,kpc large circumnuclear
ring of star formation (CNR) embracing a small bar plus 
leading arms. Those structures are P$\alpha$ emitters, but barely seen
in H$\alpha$. It presents also stellar trailing arms inside
the CNR. GEMINI and HST imagery allow the
construction of high resolution $(V-H)$ and $(J-K_s)$ color maps as
well as a $(J-K_s)$ vs. $K$ color-magnitude diagram (CMD) of this
complex region. The CNR is heavily obscured in V, but a fairly
transparent window appears in the direction of the nucleus.
Nonetheless, the nucleus presents a $(J-K_s)$ color that is redder
than the CNR. The CNR is composed of extremely young HII regions still
enshrouded in their dust cocoons. However, the nuclear $(J-K_s)$ color
can not be explained in this manner. Therefore, we propose the
contribution of C-Stars as the most feasible mechanism for explaining
the colors. If the nuclear stellar population is comparable to that of
the LMC bar, 500 C-stars and 25000 AGB O-rich stars inside 50\,pc may
reproduce the observed colors.  C-Stars release enriched material to
the nuclear environment, probably fueling the central engine of this
Sy\,2 galaxy during the lifetime of stars with masses between
2\,M$_{\odot}$ $<$ M$_{C-star}$ $<$ 6\,M$_{\odot}$ (C-star phase). The
ejected material that remains trapped in the central potential might
also explain the systematically observed increased strength of the
optical CN-bands in Sy\,2 galaxies and is consistent with the
significant contribution of intermediate age stars to the optical
continuum of low luminosity AGNs.

\end{abstract}

\keywords{Galaxies: active, nuclei, stellar content, ISM,
individual (NGC\,1241), photometry}


\section{INTRODUCTION}

NGC\,1241 is a Seyfert\,2 galaxy \citep{veron98} which presents a
complex morphology in the innermost 1.5\,kpc. P$\alpha$ imagery shows
the presence of an emitting circumnuclear ring of star formation (CNR)
with a brightness peak at a radius of $710\pm80$\,pc. It also shows a
0.3\,kpc long bar accompanied by an $m=2$ leading arm both emitting in
P$\alpha$ and centered on the nucleus. Apparently, they do not have
associated absorption features as it might be expected (Reagan \&
Mulchaey 1999, hereafter RM99). The $J$ and $K_s$ images reveal that
the CNR is mounted on a smooth inclined disk with approximately
elliptical isophotes of varying position angle. The major to minor
axis ratio of the outermost isophotes in the $J$ and $K_s$ bands
reveals a disk inclination of 52\arcdeg, consistent with the value
given by Tully (1988) for the large scale disk. Finally, the $K_s$
image shows the presence of a trailing arm ending at the CNR and
centered on the nucleus. These structures (Figure\,1) have been
kinematically studied by D\'{\i}az et al. (2003).
RM99 have found that the $(V-H)$ color map to the southwest of the
line of nodes of NGC\,1241 is redder than in the northeast area, and
reveals an overall dusty morphology consistent with an inclined ring
with a color excess of 1.1\,mag, whose southwest side is the nearest
one. According to these authors, assuming that the dust scale height
is small relative to the scale height of the stars, and that the plane
of the dust is inclined with respect to the bulge stars, dust
absorption might not affect the color of the bulge near the
nucleus. As we will see later, the nucleus of NGC\,1241 is relatively
free of absorption when compared to the color excess of the ring. To
reinforce this picture, none of the absorption features normally
expected near emitting bars are evident near the 300\,pc long bar
found by D\'{\i}az et al.  (2003). These authors have also shown that
the ring of dust found by RM99 coincides with the CNR (Figure\,2).

In this paper we examine GEMINI (+QUIRC+Hokupa) near-infrared images
with pixel-photometry. We detected the presence of an azimuthally
symmetric nuclear $(J-K_s)$ color excess with respect to the CNR
($0.82<(J-K_s)<1.15$) which would not be easily explainable in terms
of dust absorption according to the RM99 results
abovementioned. Moreover the $(V-H)$ color map does not show azimutal
symmetry as the ($J-K_s$) one does.
The properties of the $(V-H)$ and ($J-K_s$) in the inner 2\,kpc are
analyzed in terms of the models of Witt, Thronson \& Capuano (1992,
WTC92) ({\it Dust and transfer of Stellar Radiation within Galaxies}),
and stellar population synthesis from 2-MASS NIR color magnitude
diagrams of bar fields in the Large Magellanic Cloud (LMC), where
individual stars are resolved at M$_V\leq-3$ (Nikolaev \& Weinberg,
2000).

The next section of this paper outlines how our observations were made
and discuss the homogenization of Hubble Space Telescope (HST) and GEMINI photometry. In
Section\,3 our results are discussed and the final remarks are given
in the final section.

\section{OBSERVATIONS AND METHODS}

On September 30, 2000, we used the Quick Start service of the Gemini
North 8.1\,m telescope for NIR imaging using Hokupa'a natural guide
star and curvature-sensing adaptive optics system. The latter feeds
the dedicated Quick NIR camera (QUIRC) fitted with a $1024\times1024$
HgCdTe array sensitive to 1-2.5\,$\micron$ radiation providing a final
scale of 0.0197\arcsec\,pixel$^{-1}$. Standard data reduction
procedures were applied to the images. The achieved full width at half
maximum (FWHM) of the Gemini+Hokupa'a system was about 0.4\arcsec\, in
the $J$ band and about 0.3\arcsec\, in the $K_s$ band both measured on the point spread function
of a field star, and estimated on the target galaxy. Image
deconvolution was not applied at this stage because of the photometric
uncertainties that could be introduced by the methods that are
commonly used.  This resolution was
good enough to compare the Gemini images with the existing HST-NICMOS3 data (Figure 1).

\paragraph{Photometry.} 
HST imagery with F160W and F606W filters and its calibration have been
discussed by RM99. Essentially, the relative fluxes of these frames at
each position are given. A transformation was performed in order to
match colors derived from the HST fluxes $f_{F606W}$ and $f_{F160W}$
to the standard color system of our observations.
For filter F606W, we obtained $m_{F606W}$ from the flux $f_{F606W}$ in
the Vega-mag system according to
\cite{bedinetal05}. Coefficients for the transformation
of $m_{F606W}$ into $V$ of Vega-mag system are provided by
\cite{holtzmanetal95}. $V$ results $\approx0.1$\,mag
brighter than $m_{F606W}$ in agreement with previous
transformation by \cite{malkanetal95} who determined that $V$
would be 0.1\,mag to 0.2\,mag brighter than $m_{F606W}$.
For filter F160W, we calculated $m_{F160W}$ according to
\cite{stephensetal00}. These authors follow two different
procedures to transform $m_{F160W}$ into $H$, each providing
slightly different results. To be coherent with the transformation of 
$m_{F606W}$ into $V$, we adopted the procedure based on the Vega-mag
system.

Using homogeneous colors, we carried out pixel photometry to
ascertain whether the morphology seen in $K_s$ band images is
differentially affected by the presence of dust. After separating
all the pixels to the northeast from those to the southwest of the
major axis, we integrated  the $K_s$ brightness and $(J-K_s)$ and
$(V-H)$ colors on half-rings of variable radii and plotted them
against the de-projected radius (Figure 3).

Witt, Thronson \& Capuano (1992, hereafter WTC92) models of {\it Dust
and Transfer of Stellar Radiation within Galaxies} have been used to
disentangle color properties due to dust from those due to stellar
population effects. WTC92 models include the effect of light
scattering by dust. Four of the models presented by these authors
constitute plausible scenarios for the region under study: 1) {\it The
Dusty Galaxy}, which considers dust and stars equally distributed
within a sphere. 2) {\it The Cloudy Galaxy}, which considers the
sphere occupied by stars to be larger than that occupied by dust. 3)
{\it The Starburst Galaxy}, in which also stars occupy a larger sphere
than the dust, but follow an $r^{-6}$ distribution. 4) {\it The Dusty
Galactic Nucleus}, in which a sphere of stars is enshrouded in a
cocoon of dust.

Errors in colors for each one of the subsystems quoted in Figure\,3
are directly obtained from fluctuations in pixel photometry and
propagated to the derived quantities according to the transformation
equations of the photometric system.

\section{RESULTS \& DISCUSSION}

Figure\,3 shows that the $(J-K_s)$ reddening increases inwards, as
well as a remarkably similar radial behavior on both sides of the line
of nodes. It also shows a plateau at $(J-K_s)\approx0.8$\,mag at the
position of the CNR, with a maximum of about 1.15 mag in the
nucleus. The $K_s$ band integrated profile is also symmetric. On the
other hand, the upper panel of Figure\,3 shows how dramatically
different are both sides of the circumnuclear ring in the ($V-H$)
color, with an excess $E(V-H)\approx1.0$ in the southwest side with
respect to the northeast one, besides a global mean color excess
$<E(V-H)>\approx0.90$ of the CNR with respect to the disk. The nucleus
is remarkable too, as its ($V-H$) color is similar to that of the disk
outside the CNR, indicating the presence of a transparent window in
that direction as suggested by RM99. This conclusion is corroborated by
the detection of nuclear H$\alpha$ emission (the galaxy is a Sy\,2)
together with P$\alpha$ emission, while the CNR is observed in
P$\alpha$, but obscured in H$\alpha$.

Colors in  Table\,1 can be matched to stellar spectral types using
Pickles' (1998) stellar library. The results are quoted in
Table\,2.
We first note that the ($V-H$) colors of all substructures in Table\,1
correspond to younger spectral types than it would be indicated by the
$(J-K_s)$ color, in spite of the stronger reddening in the $V$ band.

\paragraph{The disk.} Its $(J-K_s)$ color is similar to the color in
the foreground of the LMC fields studied by Nikolaev \& Weinberg
(2000) and to the Sagittarius comparison fields studied by Cole
(2001), both obtained from the Two Micron All Sky Survey (2MASS) data.
We obtained similar results by integrating six Milky Way fields
around the LMC, as we will show in the results. Nevertheless, the
disk ($V-H$) color corresponds to a B9V-A2V stellar population, which
leads us to think that star formation occurred not only in the CNR
but also reached the inner disk within its innermost 2\,kpc.

\paragraph{Circumnuclear Ring (CNR).}

Assuming that the difference in the observed ($J-K_s$) between the far
side of the CNR and the disk is caused by extinction, we obtain
$E_{fs_o}(J-K_s)\approx0.30$\,mag and WTC92 {\it Dusty Galaxy} model
leads to an extinction solution with $\tau_V=6.0$ and scattered light
contributing 45$\%$ in $V$ band. This model provides theoretical color
excesses for the CNR far side amounting to
$E_{fs_t}(J-K_s)=0.30$\,mag and $E_{fs_t}(V-H)\approx0.7$, coherent with
the values presented in Table\,1. This source of absorption, probably
diffuse, is intrinsic to the CNR and different from that discussed by
\cite{regan99}, which produces the difference in ($V-H$) between
the CNR near and far sides, attributed to the dusty one-arm clearly
seen in HST F606W filter image by \cite{regan99}. Furthermore, within
the CNR one should add the H$\alpha$ absorbing dust cocoons associated
to each one of the P$\alpha$ emitting blobs (Figure\,1).

\paragraph{Nuclear colors, dust or stellar population?}

Figure 4 shows the NIR color-magnitude diagram $(J-K_s)$ vs. $K_s$ for
0.1\arcsec\, rebinned pixels in the central region of NGC\,1241. The
figure presents a color excess $E_N(J-K_s)\approx0.2$ of the nucleus
with respect to the CNR in a different manner (cf. Figure\,3). It is
not possible to obtain a WTC92 model fitting of both the very red
($J-K_s$) color excess and the very blue nuclear ($V-H$). The
reddening arrow in Figure\,4 points to the same direction where the
nuclear $(J-K_s)$ at the top of the CMD bends. Therefore, it suggests
that dust may affect in a rather subtle way the nuclear $(J-K_s)$
without influencing ($V-H$).  Our solution to this rather tricky issue
is that Carbon stars are natural candidates to explain the infrared
excess. Considerations on the LMC infrared Color-Magnitude diagram (CMD) of
Nikolaev \& Weinberg (2000), lead us to propose these substantial
contribution from Carbon stars to the nuclear stellar population.
Following a procedure similar to that of Nikolaev \& Weinberg (2000),
we have used the 2MASS All-Sky Point Source Catalog Statistics Service
facility at {\it http://irsa.ipac.caltech.edu/applications/Stats/} to
obtain an integrated color-magnitude synthesis of the LMC bar and we
have compared it to the NGC\,1241 nucleus.

The 2MASS service provides CMD diagrams and integrated light
photometry in $J$, $H$ and $K$ inside user selected circular regions.
We choose five circular fields along the LMC bar and 2 foreground
regions per bar field, one located 10 degrees North and another 10
degrees South of the bar. All regions and fields were selected with
30' radius. Extractions from the 2MASS Point Source Catalog were
performed. Figure\,5 shows the CMD diagram corresponding to the center
of the LMC bar and to the corresponding comparison field. Then, for
each of the five regions the North and South foreground fields were
averaged, and the mean brightness subtracted from the corresponding
region to correct for foreground contamination. The final integrated
surface magnitudes were obtained by flux averaging the five background
corrected integrated surface magnitudes of the regions. The results
were: $<J>\,=20.38\pm0.07$\,mag/arcsec$^2$,
$<H>\,=19.59\pm0.08$\,mag/arcsec$^2$,
$<K>\,=19.38\pm0.07$\,mag/arcsec$^2$, $<(J-K)>\,=1.00\pm0.10$. The
$<(J-K)>$ color closely agrees with the color of the nucleus of
NGC\,1241. A similar color was obtained by Cole (2001) for the
Sagittarius Dwarf galaxy, after correction for our Galaxy
contamination.

\section{Final Remarks}

We have discussed two dimensional photometry of the central 2\,kpc
of the Sy\,2 galaxy NGC 1241, where a circumnuclear ring of star
formation and the nucleus present peculiar colors when compared to
the underlying disk. HST and GEMINI imagery have been reduced to a
uniform photometric system in order to allow the study of the
photometric properties of these subsystems.
While the dust arm produces the reddening of the CNR near side with
respect to the CNR far side, we propose that an additional source of
diffuse dust obscures uniformly the CNR, thus producing a global
reddening of the CNR compared to the underlying disk. Inside the CNR,
there are cocoons of dust associated to the P${\alpha}$ emitting
condensations.

Finally, the very red $(J-K_S)$ color of the nuclear region together
with the surprising transparency of this region in $(V-H)$, led us to
propose a CMD for the nucleus similar to that of the LMC bar.  C-stars
can in fact redden significantly the integrated colors at unresolved
scales, a situation similar to that we are facing in the nuclear
region of NGC\,1241. Carbon stars and asymptotic giant branch
oxygen-rich stars evolve rapidly ($t < 3\times10^4$\,yr) and eject
considerable amounts of dust and gas with velocities low enough
($V_{gas} < 100$\,km s$^{-1}$) to be trapped by the gravitational
potential barrier of the central mass concentration
($M_{kepler}\sim10^9$\,M$_{\odot}$, $r<300$\,pc). The 500
[6\,M$\odot$] C-stars and $2.5\times10^4$ asymptotic giant branch
O-rich stars (according to the LMC bar proportion) inside a radius of
about 50\,pc that are necessary to explain the nuclear colors, would
release material that, gravitationally bounded, could amount to
between $10^{-2}$ and $10^{-1}$ M$_{\odot}$\,yr$^{-1}$ of fuel for the
central engine. The intra-nucleus medium contamination may last during
the lifetime of stars with masses
2\,M$_{\odot}<$\,\,M$_{C-Stars}<6\,$M$_{\odot}$. This scenario may
also explain the systematical  increase of the strength of the
optical CN-bands observed in the stellar populations of Sy\,2 galaxy nuclei
(e.g. Gu et al. 2001), and the significant contribution of
intermediate age stars to the optical continuum of low luminosity AGNs
(e.g. Gonzalez-Delgado 2004).

\section{Acknowledgements}

HD thanks the brazilian institutions CNPq and
CAPES. RD thanks the support from Evencio Mediavilla and Romano Corradi.
This research is also partially supported by brazilian grants
MEGALIT/Millennium, and the argentinean Agencia C\'ordoba Ciencia. 
JFS thanks to the FAPEMIG Foundation (Minas 
Gerais, Brazil). 
The 2MASS project is a collaboration between the University
of Massachusetts and the IPAC (JPL/Caltech). 
The Gemini 8-meter telescopes is an international partnership
managed by AURA, Inc. under a cooperative agreement with the
NSF (USA), PPARC (UK), NRC (Canada), CONICET (Argentina), ARC (Australia),
CNPq (Brasil) and CONICYT (Chile). The NASA/ESA Hubble Space
Telescope is operated by AURA under NASA contract NAS 5-26555.




\clearpage

\begin{figure}
\includegraphics{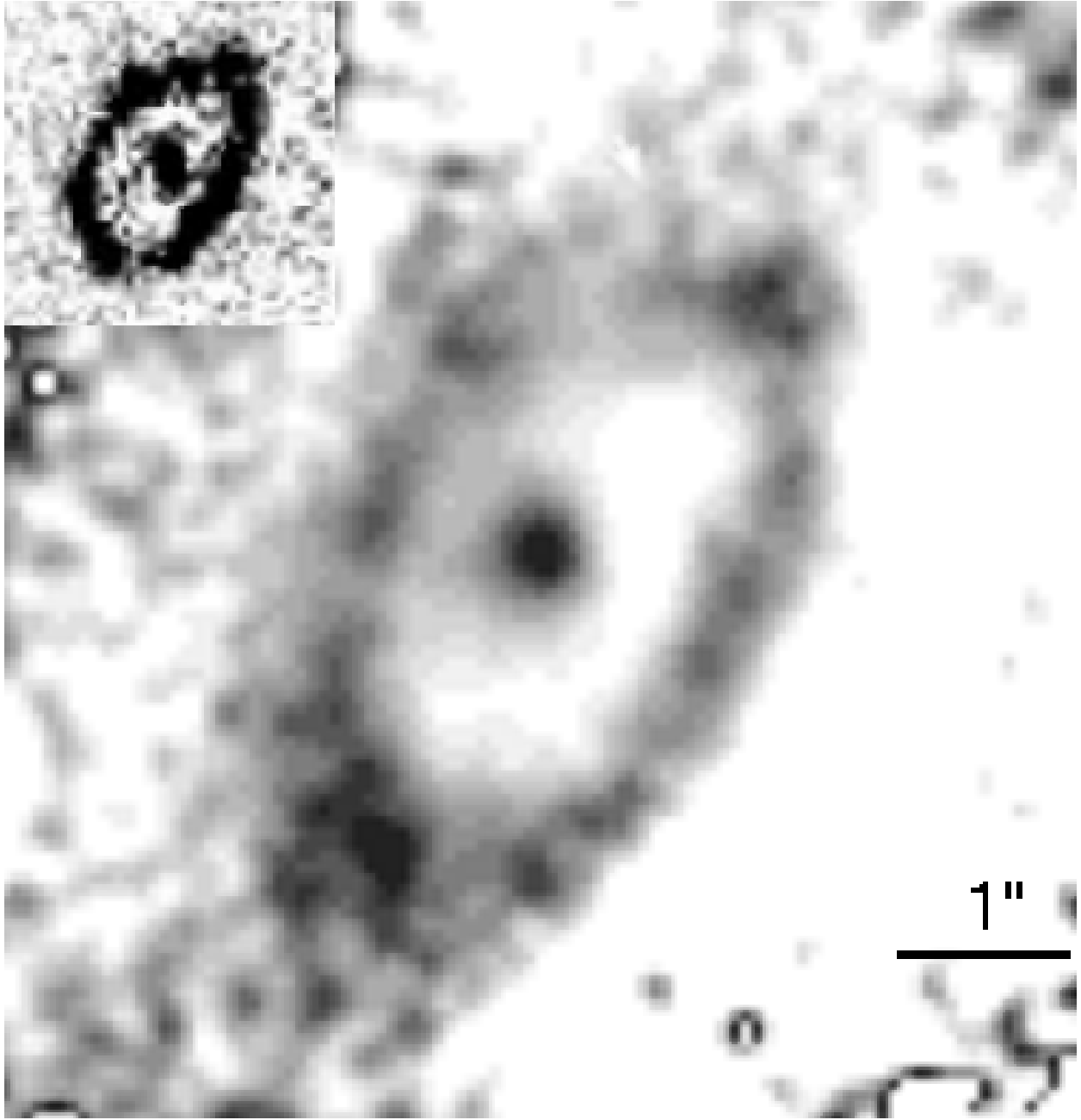} \vspace{19 cm}
\caption{Gemini $J-K_s$ 
differential color map of the central $2\times2$\,kpc region of
NGC\,1241. North is at the top and East to the left. The darkest areas
represent those regions which more strongly depart from a smooth color
distribution. The inset shows the HST Pa\,$\alpha$ image, after
continuum subtraction using the $H$-band. Note the bar-like
central structure and the faint leading spiral in the Pa\,$\alpha$
emission (see D\'{\i}az et al. 2003).}
\label{figure1}
\end{figure}

\clearpage

\begin{figure}
\includegraphics{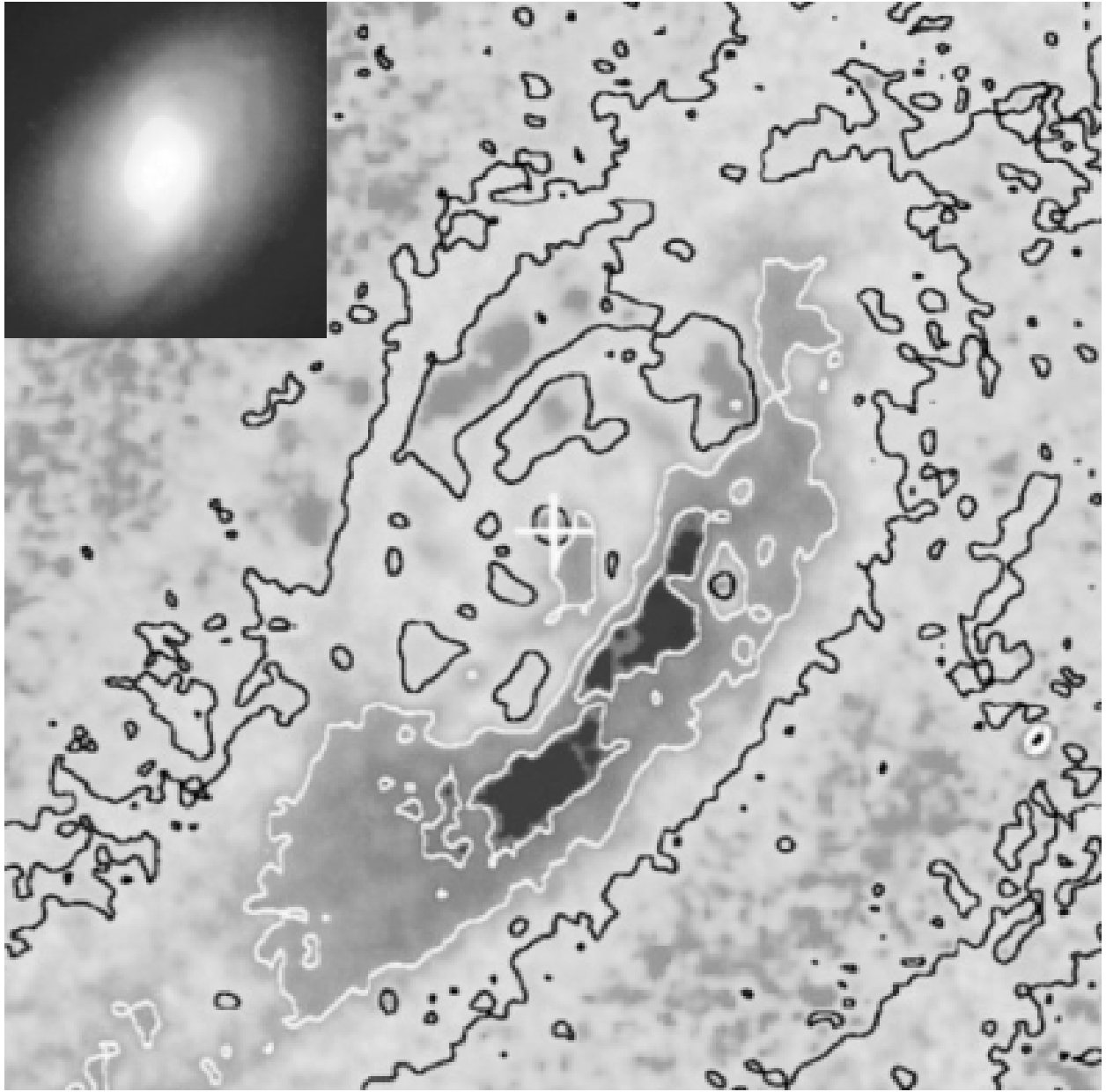} \vspace{19 cm}
\caption{Color map made by combining
the NICMOS 1.6$\micron$ image with a HST-WFPC2 F606W image (Regan
\& Mulchaey 1999). The contours of the F606W-F160W red color
excesses (compared with the background galaxy) are at magnitude
0.3, 0.6, 0.9 and 1.2.  The field shows the central $2\times2$\,kpc 
region of NGC\,1241. North is at the top and East to the left. 
The inset shows the $H$-band (F160W) image.} \label{figure2}
\end{figure}

\clearpage

\begin{figure}
\includegraphics{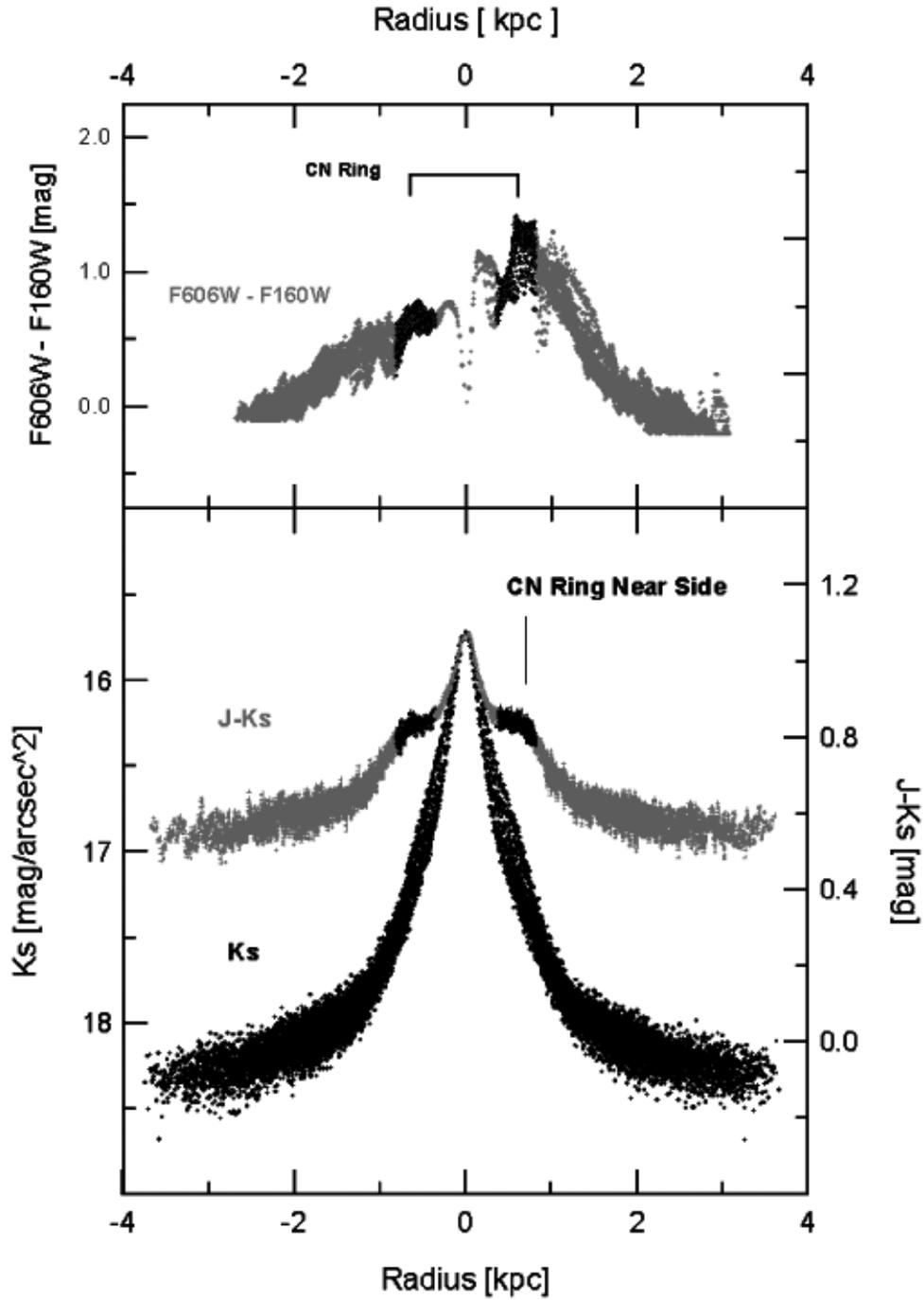} \vspace{19 cm} \caption{$V-H$ (upper panel), $K_s$ and
$J-K_s$ band photometric values for 0.1\arcsec\, rebinned pixels
plotted against the de-projected radius. Pixels to the northeast are
separated from those to the southwest of the major axis (positive
radii). The pixels in the radial range corresponding to the CNR were
plotted as black points in the color profiles. Note the increasing
reddening inwards, and the remarkably similar radial behavior of 
the $J-K_s$ color profile on both sides of the line of nodes.}
\label{figure3}
\end{figure}

\clearpage

\begin{figure}
\includegraphics{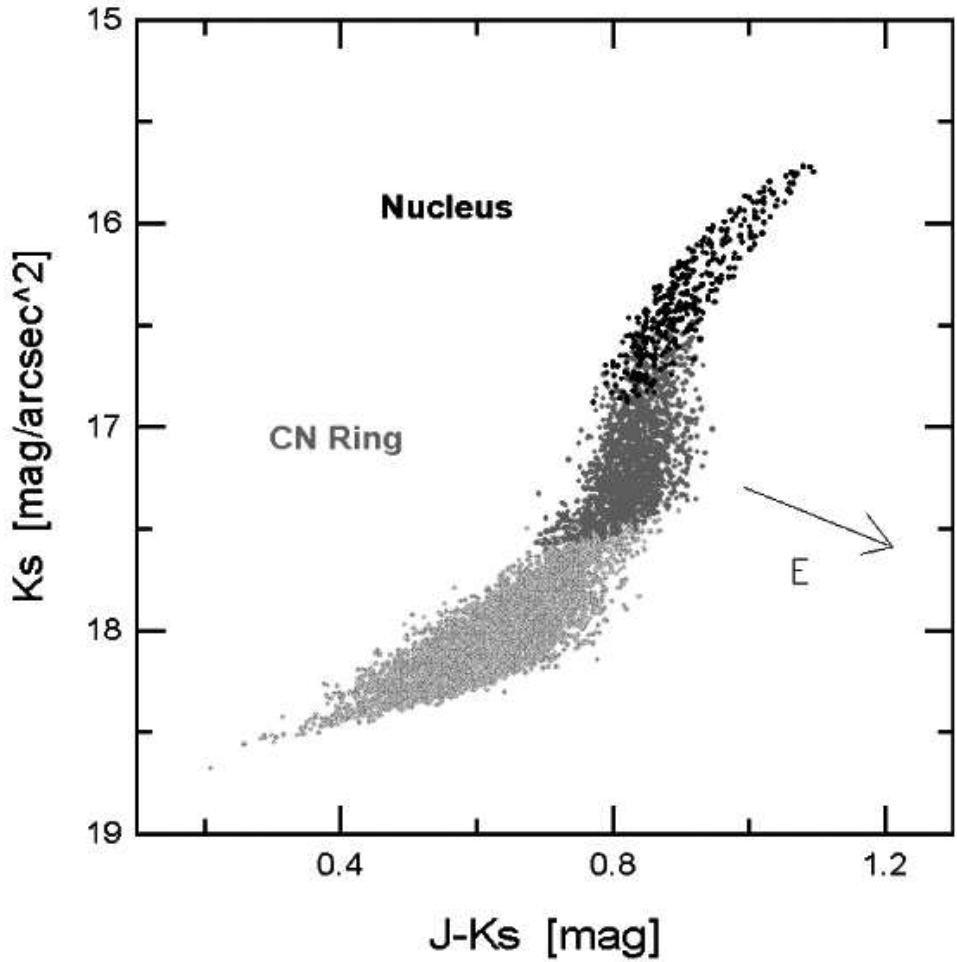}
\vspace{19 cm} \caption{NIR color-magnitude plot for 0.1\arcsec\,
rebinned pixels. The radial positions corresponding to the nuclear region 
and the circumnuclear ring, and the reddening
direction (representing $E(J-K_s)\approx0.2$) have been noted.} \label{figure4}
\end{figure}

\clearpage

\begin{figure}
\includegraphics{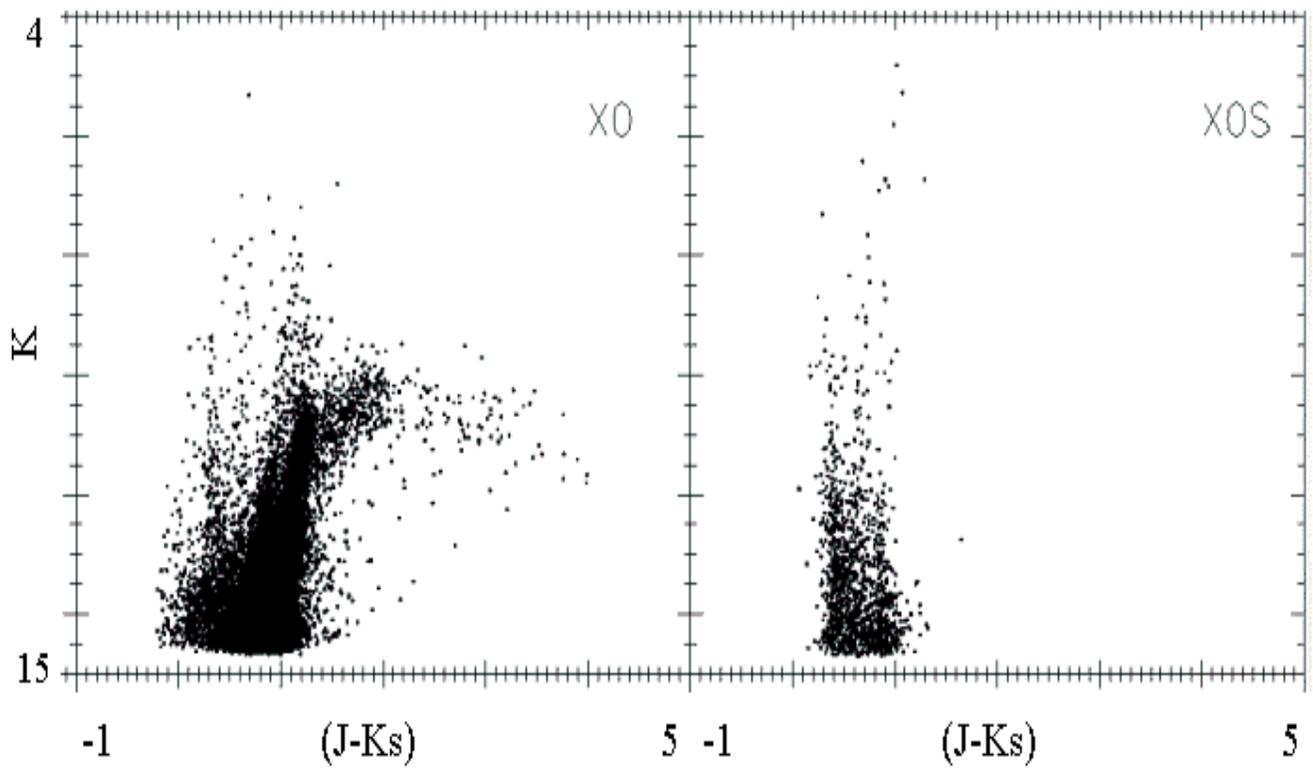} 
\vspace{19 cm}
\caption{Color-Magnitude diagram of the LMC bar center (left), 
and a Milky Way field (right) located 10 degrees North of the bar.} 
\label{figure5}
\end{figure}


\begin{deluxetable}{lcc}
\tablewidth{30pc} \tablecaption{NGC\,1241 Colors and relative fluxes} \label{tab0}
\tablehead{ \colhead{Structure} & \colhead{$(V-H)$}& \colhead{\bf $(J-K_s)$}\,\tablenotemark{a}\\
\colhead{} &\colhead{($m_{F606W}-m_{F160W}$)} &
\colhead{($m_{110W}-m_{F222W}$)}}
\startdata
\\
Disk (outwards CNR)&-0.1$\pm$0.2& {\bf 0.55$\pm$0.1}\\
 & {\bf 0.0$\pm$0.2}&\\
\\
CNR (near side)&1.5$\pm$0.2& {\bf 0.82$\pm$0.06}\\
 &{\bf 1.35$\pm$0.2}&\\
\\
CNR (far side)&0.61$\pm$0.1& {\bf 0.82$\pm$0.06}\\
&{\bf 0.6$\pm$0.05}\\
\\
Nucleus&0.0$\pm$0.1  & {\bf 1.05$\pm$0.03}\\
&{\bf 0.0$\pm$0.1}&\\
\\
\enddata
\tablenotetext{a}{\,\,$J$ and $K_s$ are standard colors observed with GEMINI.
$m_{F606W}$ and $m_{F160W}$ are derived from HST fluxes. Observed quantities are
in bold.}
\end{deluxetable}

\clearpage

\tablecaption{Stellar spectral types in Pickles' (1998) stellar library
that match the colors in Table 1}
\begin{tabular}{|c|c|c|}
  \hline
  Structure & $(V-H)$ & $(J-K_s)$ \\
  \hline
  disk (outwards CNR)&B9V-A2V & K0-K2 \\
  CNR (near side) &G2V-G5V& M2-M4 \\
  CNR (far side)&F0V-F2V & M1-M2 \\
  Nucleus& A3V-A5V& M5-M6 \\ \hline
\end{tabular}


\begin{thebibliography}{}

\bibitem[Bedin et al.(2005)]{bedinetal05} Bedin, L.R., et al. 2005, MNRAS, 357, 1038.

\bibitem[Cole(2001)]{cole03} Cole, A.A.,2001, ApJ, 559, L17.

\bibitem[D\'{\i}az et al.(2003)]{diaz03} D\'{\i}az, R.J., Dottori, H.,
Vera-Villamizar, N., Carranza, G. 2003, ApJ, 597, 860.

\bibitem[Gonzalez-Delgado(2004)]{gonzalez04} Gonzalez-Delgado,
R., 2004, in The Interplay among Black Holes, Stars and ISM in
Galactic Nuclei, L. Ho \& H.
Schmitt, eds., Cambridge University Press 2004, p. 137-140.

\bibitem[Gu et al.(2001)]{gu01} Gu, Q., Huang, J., de Diego, J.,
Dultzin-Hacyan, D., Lei, S. \& Ben\'{\i}tez, E. 2001, A\&A, 374, 932.

\bibitem[Holtzman et al.(1995)]{holtzmanetal95} Holtzman, J.A.,
Burrows, C.J., Casertano, S., Hester, J.J., Trauger, J.T., Watson,
A.M., Worthey, G. 1995, PASP, 107, 1065.

\bibitem[Malkan, Gorjian \& Tam(1995)]{malkanetal95} Malkan, M.A.,
Gorjian, V., Tam, R. 1995, ApJSS, 117, 25.

\bibitem[Nikolaev \& Weimberg(2000)]{nikolaev2000} Nikolaev, S., \&
Weinberg, M. 2000, ApJ, 542,804.

\bibitem[Pickles(1998)]{pickles98} Pickles, A. J. 1998, PASP, 110, 863.

\bibitem[Regan \& Mulchaey(1999)]{regan99} Regan, M.W., \& Mulchaey,
J.S. 1999, AJ, 117, 2676.

\bibitem[V\'eron-Cetty et al.(1998)]{veron98} V\'eron-Cetty, M., V\'eron, P. 1998, ESO Sci. Report: A Catalogue of
Quasars and Active Nuclei, 8th Edition.

\bibitem[Stephens et al.(2000)]{stephensetal00} Stephens, A.W.,
Frogel, J.A, Ortolani, S., Davies, R., Jablonka, P., Renzini, A.,
Rich, R. M. 2000, AJ, 119, 419.

\bibitem [Witt et al.(1992)]{witt92} Witt, A.N., Thronson Jr., H.A., Capuano Jr., J.M 1992, ApJ, 393, 611.


\end{thebibliography}
\end{document}